# The mean light curves of the Mira-type stars in the H- and K-bands

## L. S. Kudashkina

Department "Mathematics, Physics and Astronomy"
Odessa National Maritime University, Odessa, Ukraine
kudals04@mail.ru

Abstract. For nine Mira-type stars (o Cet, R Leo, S Car, U Her, X Oph, R Aql, RR Aql, S Ori, S Scl,) and one semi-regular star ($L_2$ Pup), the mean light curves have been obtained. The initial values of the brightness (observations) were fitted by a trigonometric polynomial of the statistically optimal order.

The Fourier coefficients and the additional parameters (degree of the trigonometric polynomial, amplitude of the brightness, the maximal slope of ascending and descending branch, semi-amplitudes and initial epochs for the brightness maximum (minimum magnitude) of each harmonic contribution, etc., were given in the tables reported earlier (Kudashkina, 2015). In this study, we have received several interesting correlations between these parameters in the bands H and K. It was particularly noted the anomalous position on the figures of the star X Oph.

The mean light curves of the investigated stars are commonly symmetric in the near infrared region (H and K).

This article is a continuation of the work on the Atlas of the mean light curves and catalogue of the characteristics of the long-period variable stars (Kudashkina, 2003; Chinarova and Andronov, 2000; Marsakova and Andronov, 1998) and these works were done in a frame of activity of the "Ukrainian Virtual Observatory" (Vavilova et al., 2011, 2012) and "Inter-Longitude Astronomy" (Andronov et al., 2003). The "Stellar Bell" (pulsating stars) part of this campaign was reported by Andronov et al. (2014).

More details on observational bases, methods of processing of observations and methods of approximation of curves can be found in Andronov (2003) and Andronov and Marsakova (2006).

For this investigation, the observational data of the near infrared bands (H and K) from the article «Infrared colours for Mira-like long-period variables found in the Hipparcos catalogue» (Whitelock et al., 2000) were used. 10 stars were processed: o Cet, R Leo, S Car, U Her, X Oph, R Aql, RR Aql, S Ori, S Scl, $L_2$ Pup.

As mentioned in the articles by Kudashkina (2015), Kudashkina and Andronov (1996), for the analysis we have used the program by Andronov (1994, 2003), which allows the use of a trigonometric polynomial fit of the statistically optimal degree $s$:

$$m(t) = a_0 - \sum_{k=1}^{s} r_k \cos(2\pi k \cdot (t - T_{0k})/P),$$

where $r_k$ are semi-amplitudes and $T_{0k}$ are initial epochs for the brightness maximum (minimum magnitude) of the wave with a period $P_k = P/k$.

The preliminary value of the period (from the General Catalogue of Variable Stars, (Samus et al. 2007-2015) was corrected using the method of differential corrections for each order $s$ of the trigonometric polynomial (Andronov, 1994, 2003).

All computed parameters of light curves are subdivided into three groups: first, fundamental (period P, amplitude $\Delta m = m_{min} - m_{max}$, asymmetry $f = \varphi_{max} - \varphi_{min}$, degree of the trigonometric polynomial $s$); second, parameters of the extremal slope of the light curve ($m_i$ and $m_d$ – the maximal slope of the incline for ascending and descending branches; $t_i$ and $t_d$ – the characteristic time of the increase of brightness by $1^m$ for ascending and descending branches); third, additional (parameters of harmonics).

The results were given in the tables reported earlier (Kudashkina, 2015). The main light curves are shown in the pictures. The mean light curves for $o$ Cet, R Aql and R Leo were given as well.

The regression lines and coefficient correlations were calculated for the next dependences:

(1) period – amplitude "lg$P$ – $\Delta$mag (H)", "lg$P$ – $\Delta$mag (K)";

(2) period – maximal slope of the incline for ascending and descending branches "lg$P$ – $m_i$ (H)", "lg$P$ – $m_i$ (K)", "lg$P$ – $m_d$ (H)", "lg$P$ – $m_d$ (K)";

(3) period – characteristic time of the increase of brightness by $1^m$ for ascending and descending branches "lg$P$ – $t_i$ (H)", "lg$P$ – $t_i$ (K)", "lg$P$ – $t_d$ (H)", "lg$P$ – $t_d$ (K)";

(4) period – semiamplitude of the first harmonic "lg$P$ – $r_1$ (H)", "lg$P$ – $r_1$ (K)";

(5) semiamplitude of the first harmonic – maximal slope of the incline for ascending and descending branches "$r_1 - m_i$ (H)", "$r_1 - m_i$ (K)", "$r_1 - m_d$ (H)", "$r_1 - m_d(K)$";

(6) semiamplitude of the first harmonic – characteristic time of the increase of brightness by $1^m$ for ascending and descending branches "$r_1 - t_i$ (H)", "$r_1 - t_i$ (K)", "$r_1 - t_d$ (H)", "$r_1 - t_d$ (K)";

(7) maximal slope of the incline for ascending branch – characteristic time of the increase of brightness by $1^m$ for descending branch "$m_i - t_d$ (H)", "$m_i - t_d$ (K)";

(8) maximal slope of the incline for descending branch – characteristic time of the increase of brightness by $1^m$ for ascending branch "$m_d - t_i$ (H)", "$m_d - t_i$ (K)".

(9) difference of the amplitudes in the H – and K – band versus period $\Delta m(H) - \Delta m(K) - $ lg $P$.

Conclusions

For (8) the coefficient of the correlation for K – band $\rho=0.92$ and $\rho/\sigma_\rho=6.9$. Meanwhile, the correlation coefficient for the H – band is only 0.24. Perhaps this is because the light curve in the K – band is more symmetrical than in the H – band.
For (1) the coefficients of the correlation are $\rho=0.55$ and $\rho/\sigma_\rho=1.9$ (H); $\rho=0.54$ and $\rho/\sigma_\rho=1.8$ (K).
For (4) the coefficients of the correlation are $\rho=0.52$ and $\rho/\sigma_\rho=1.7$ (H); $\rho=0.52$ and $\rho/\sigma_\rho=1.7$ (K).

For all other dependencies of the correlation coefficient less than 0.5 in absolute value and the ratio $\rho/\sigma_\rho$ never exceeds 2.

Ratio (8) prompted the idea to test the dependence of the difference of the amplitudes of the rays H and K on the logarithm of the period (9).
For all ten stars of the dependence (9), the correlation coefficient is small enough ($\rho=0.596$ and $\rho/\sigma_\rho=1.8$). If we exclude two stars with highly asymmetric curves (S Ori and S Scl), the picture becomes different: $\rho=0.70$ and $\rho/\sigma_\rho=2.4$ (fig. 6). However, the star X Oph is also outstanding at the picture. This object is an anomalous (small-amplitude) Mira-type star, which is also a spectral binary. Its features were discussed earlier (Andronov and Kudashkina, 2008). Also previously described features of the Mira-type stars R Aql, R Leo, U Her in the article Kudashkina and Rudnitskij (1988), in particular, we compared the behavior of the stars in the visual range and in the maser of the $H_2O$ line.

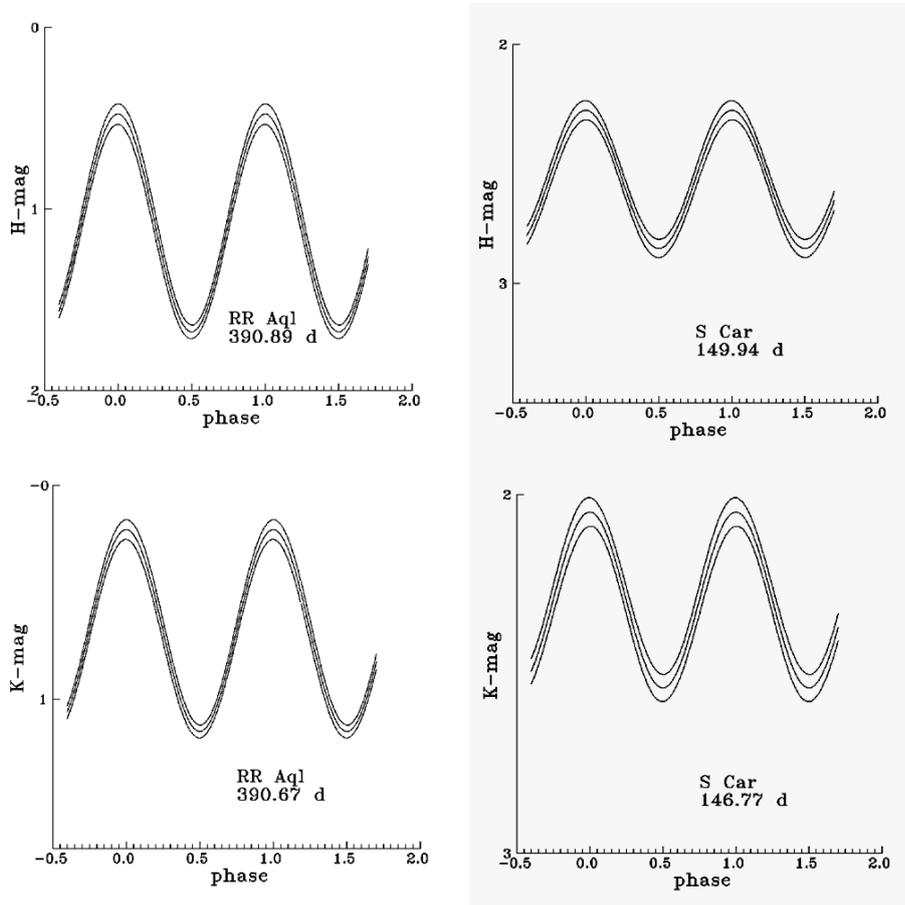

Fig. 1. The mean light curves of the Mira-type stars in H and K bands. The best fit and the ±1σ error corridor are shown.

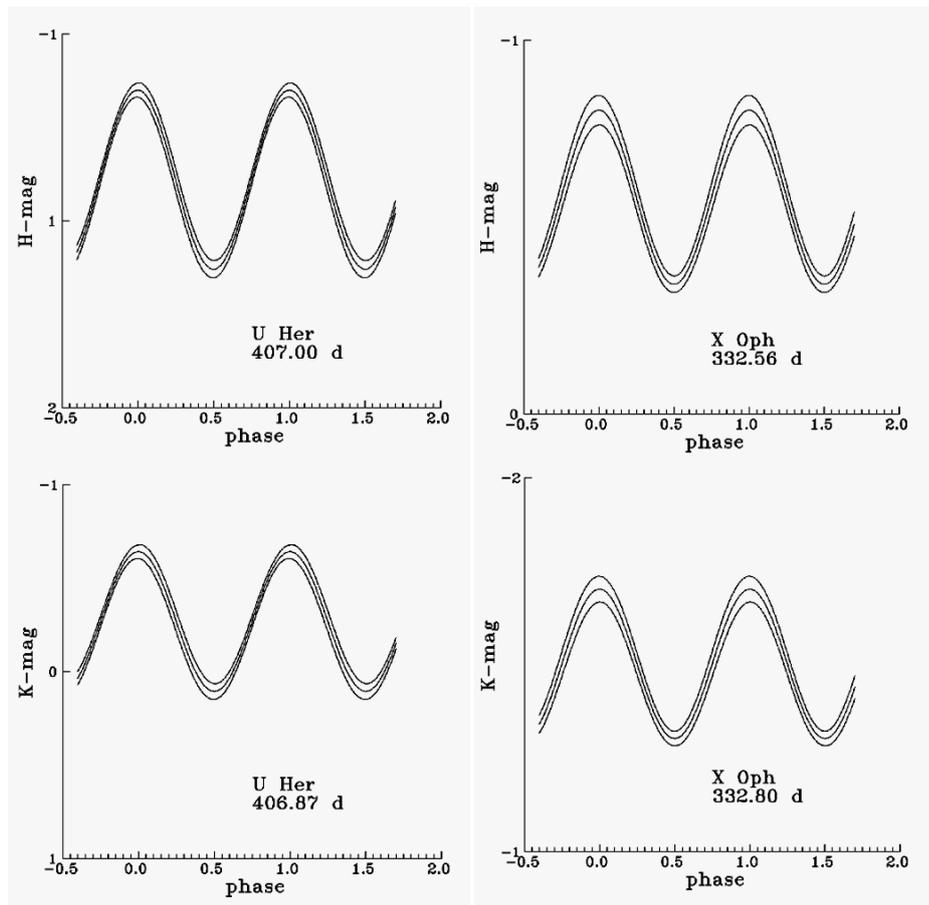

Fig. 2. The mean light curves of the Mira-type stars in H and K bands. The best fit and the ±1σ error corridor are shown.

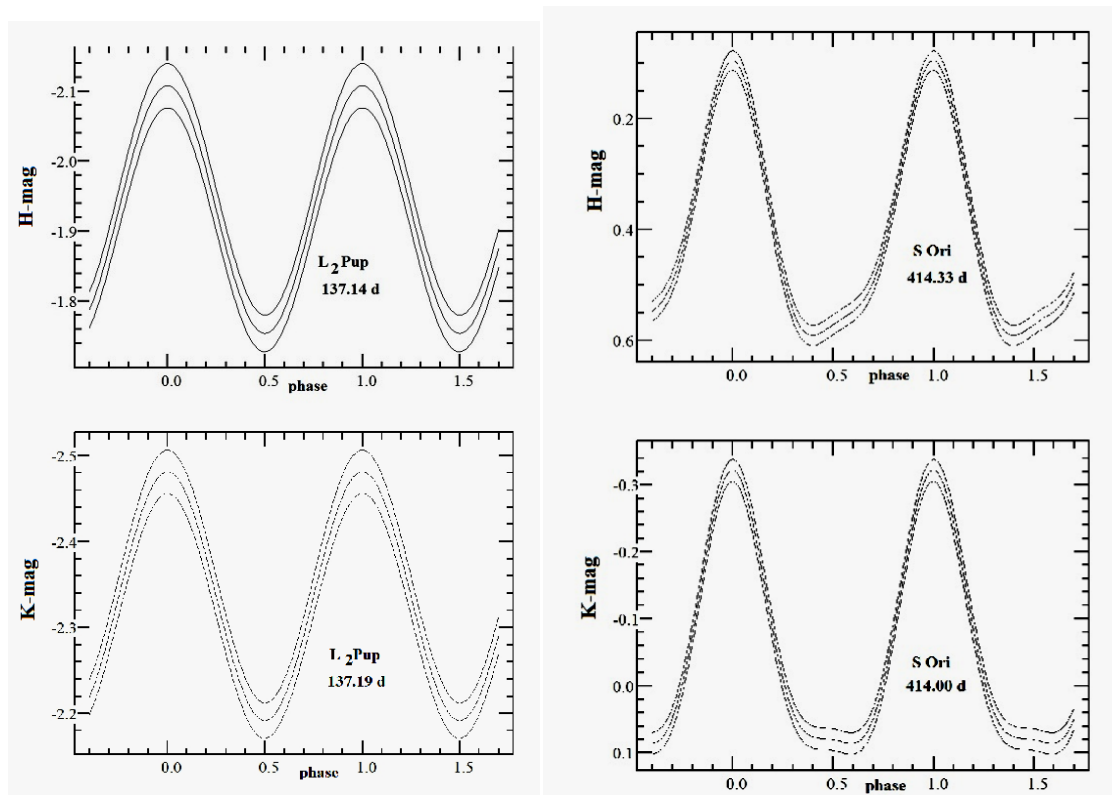

Fig. 3. The mean light curves of the Mira-type stars in H and K bands. The best fit and the ±1σ error corridor are shown.

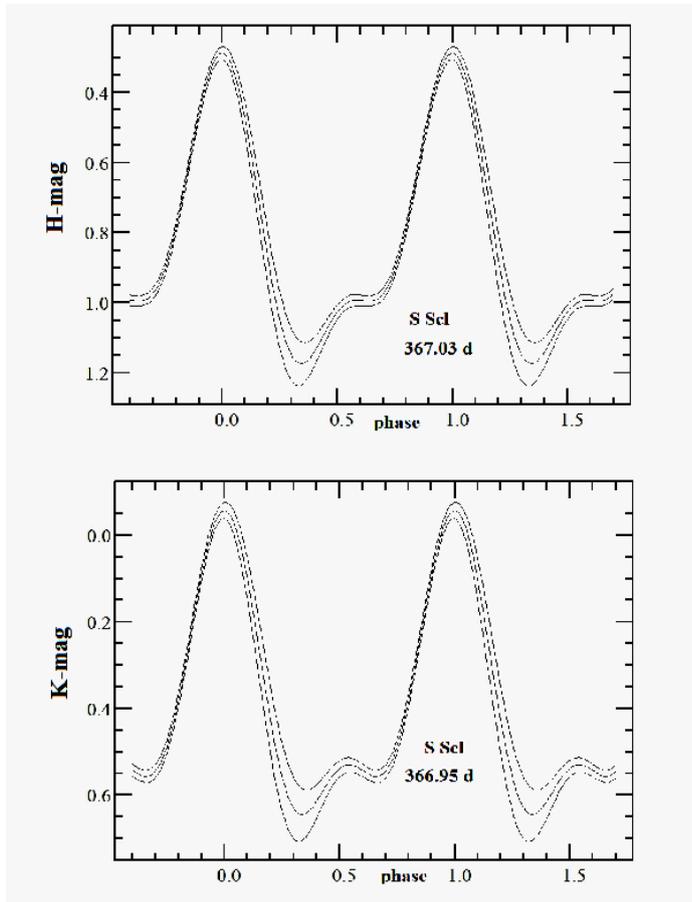

Fig. 4. The mean light curves of the Mira-type stars in H and K bands. The best fit and the ±1σ error corridor are shown.

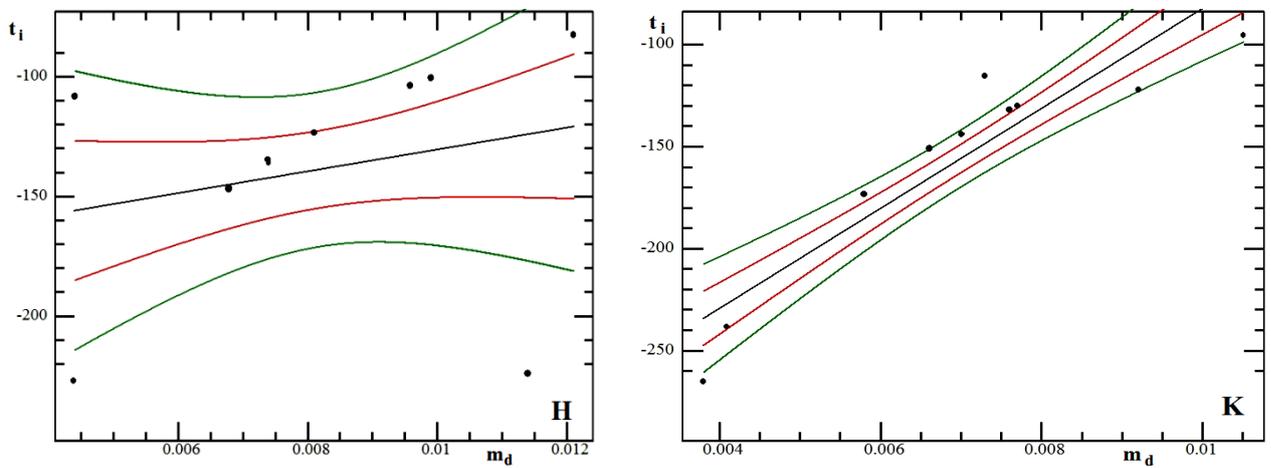

Fig. 5. The dependencies between mutually inverse characteristics of the asymmetry of the branches. Index "i" corresponds to ascending branch, index "d" corresponds to the descending branch.

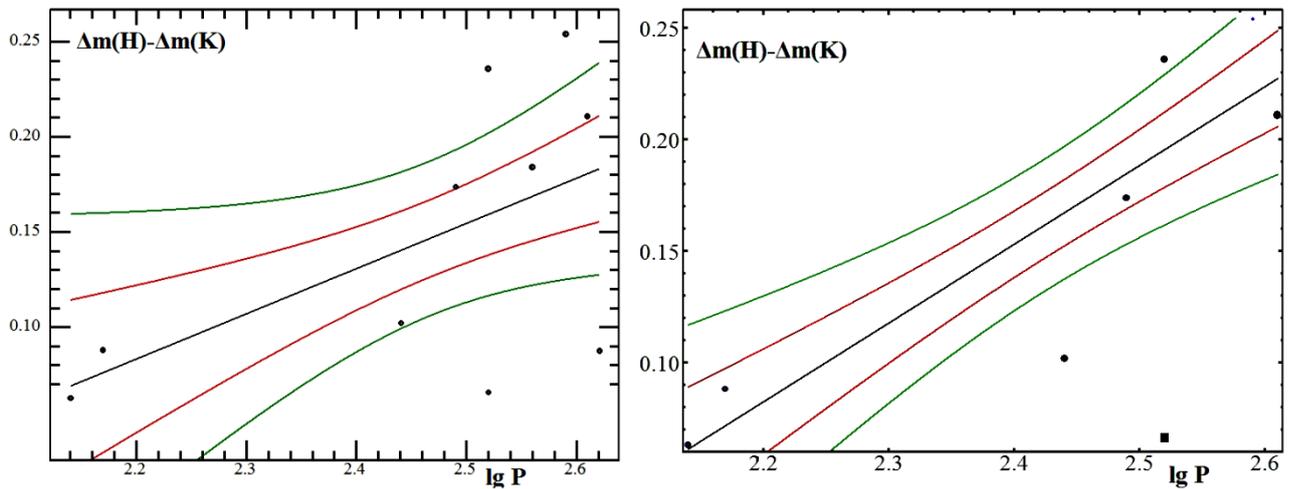

Fig. 6. The dependence between *Δm(H)* - *Δm(K)* and *lg P*. For the left dependence, ρ=0,596 and ρ/σ$_ρ$=1,8. For the right dependence ρ=0.88 and ρ/σ$_ρ$=4.2. At the bottom, the filled square shows the star X Oph, which was also excluded from the calculation.

All the figures were constructed with the program MCV (Andronov and Baklanov, 2004).